\documentclass{PoS}

\title{A model for jets of low-mass microquasars}

\ShortTitle{A model for jets of LMMQs}

\author{{Gabriela S. Vila}\\ %%\thanks{A footnote may follow.}\\
        Instituto Argentino de Radioastronom\'ia (IAR-CONICET)\\
        C.C. N$^\circ$ 5, C.P. 1894, Villa Elisa, Buenos Aires, Argentina\\
        E-mail: \email{gvila@iar-conicet.gov.ar}}
        
\author{\speaker{Gustavo E. Romero}\\
        Instituto Argentino de Radioastronom\'ia (IAR-CONICET) \\
        C.C. N$^\circ$ 5, C.P. 1894, Villa Elisa, Buenos Aires, Argentina \\
        Facultad de Ciencias Astron\'omicas y Geof\'isicas (FCAG, UNLP)\\
        Paseo del Bosque s/n, La Plata, Buenos Aires, Argentina\\
        E-mail: \email{romero@iar-conicet.gov.ar}}

\abstract{In this work we present a new jet model for the non-thermal broadband emission of low-mass microquasars. We calculate the contribution of relativistic particles, primary electrons and protons as well as secondary muons, 
charged pions and electron-positron pairs, to the electromagnetic spectrum of the sources.  The distribution in energy 
of all particle species is obtained for an extended, inhomogeneous region. We include detailed analysis of particle 
energy losses, injection, decay and escape from the acceleration zone. We also calculate absorption effects due to 
photon-photon annihilation.  As an application, we consider the case of XTE J1118+480, a well-known low-mass X-ray 
binary in the galactic halo, and we present predictions about its high-energy radiation.}

\FullConference{25th Texas Symposium on Relativistic Astrophysics - TEXAS 2010\\
		December 06-10, 2010\\
		Heidelberg, Germany}

\begin{document}

\section{Introduction}

Microquasars are X-ray binaries that produce jets, whose presence is inferred due to their characteristic synchrotron radio spectrum. The detection of non-thermal radiation reveals that microquasar jets are able to accelerate particles (at least electrons) up to relativistic energies. Four microquasars have been detected in gamma-rays; these very high-energy photons are thought to be emitted at the jets. The origin of the gamma-ray emission is believed to be leptonic, but it is possible that relativistic protons, or even secondary pions, muons and pairs, may contribute as well. 

In this work we present a jet model for the broadband spectrum of low-mass microquasars (LMMQs). These are systems where the companion star is old, dim and without strong winds. Since the star cannot provide dense targets, it is expected that in LMMQs any high-energy emission is produced by the interaction of the relativistic particles with the magnetic, matter and radiation fields of the same jets. 

In our model, acceleration of electrons and protons takes place in an extended, inhomogeneous region of the jets. The energy distribution of relativistic particles is calculated taking into account cooling, injection and convection. We then obtain the broadband radiation spectrum due to several interaction processes, including the contribution of secondary particles. We also assess the relevance of photon absorption through pair creation in photon-photon annihilation.

Finally, we apply the model to the broadband spectrum of the low-mass X-ray binary XTE J1118+480. This system is located at a distance $\approx 1.72$ kpc in the galactic halo (galactic latitude $b\approx 62.3^\circ$), and consists of a K-type star and an $8.5M_\odot$ black hole (Gelino et al. 2006). Although no jets have been directly imaged yet, the observed radio emission suggests the presence of outflows during the low-hard state (Fender et al. 2001). Applying our model, we fit the observed radio-to-X-rays spectrum of this source and make predictions for its high-energy emission. 

\section{Jet model}

The model described here is presented in detail in Reynoso \& Romero (2009) and Romero \& Vila (2008). Two symmetrical jets are injected at a distance \mbox{$z_0 = 50R_g$} from a black hole of mass \mbox{$M_{\rm{BH}}$;} \mbox{$R_g = GM_{\rm{BH}}/c^2$} is the gravitational radius of the black hole. The outflow advances with a bulk Lorentz factor \mbox{$\Gamma_{\rm{jet}} = 2$,} expanding laterally as a cone of half-opening angle \mbox{$\phi_{\rm{jet}}\approx 6^\circ$.} We assume that the jet terminates at \mbox{$z = z_{\rm{end}}$.} The angle between the jet axis and the line of sight is \mbox{$\theta = 30^\circ$.}

The accretion power of the black hole is a fraction of its Eddington luminosity, \mbox{$L_{\rm{accr}} = q_{\rm{accr}}L_{\rm{Edd}}$.} Following the disk-jet coupling hypothesis of Falcke \& Biermann (1995), we assume that a fraction of the accretion energy powers the jets: \mbox{$2L_{\rm{jet}} = q_{\rm{jet}}L_{\rm{accr}}$,} with $q_{\rm{jet}}=0.1$. The factor two takes into account the existence of a jet and a counterjet.  

The mean value of the magnetic field at the jet injection point, $B_0 \equiv B(z_0)$, is determined demanding equipartition between the magnetic and kinetic energy densities. For larger $z$ we parameterize the magnetic field as a power law, $B(z)=B_0(z_0/z)^{1.5}$.

Part of the jet power is converted into kinetic energy of relativistic particles through a diffusive shock acceleration mechanism. The power injected in relativistic particles is $L_{\rm{rel}} = q_{\rm{rel}}L_{\rm{jet}}$, with $q_{\rm{rel}}=0.1$. Both protons and electrons are accelerated, so $L_{\rm{rel}} = L_e + L_p$ where $L_p$ and $L_e$ are the power injected in relativistic protons and electrons, respectively. We relate these two quantities as $L_p = aL_e$, with $a\geq1$ a free parameter. 

The acceleration region is located at $z_{\rm{acc}} > z_0$. The value of $z_{\rm{acc}}$ is such that the magnetic energy density is smaller that the jet kinetic energy density at that distance from the black hole (see Komissarov et al. 2007 for a discussion on this topic). The injection function of relativistic primary protons and  electrons is assumed to be of the form

\begin{equation}
Q(E,z) = Q_0\, f(z)\,E^{-\alpha}\exp\left(-\frac{E}{E_{\rm{max}}}\right)   \quad\quad \left[Q\right]= \rm{erg}^{-1}\,\rm{cm}^{-3}\,\rm{s}^{-1}.
\label{eq:injection_function}
\end{equation}

\vspace{0.2cm}     

\noindent Here $E_{\rm{max}}(z)$ is the maximum energy that particles can achieve at fixed $z$; it is determined by the balance of the total energy  loss rate and the acceleration rate (see below). The function $f(z)$ is given by a step-like function, that is unity for $z\lesssim z_{\rm{max}}$ and becomes negligible for larger $z$.

Neutral and charged pions are created through the interaction of relativistic protons with photons and non-relativistic protons. Whereas neutral pions decay into photons, charged pions yield neutrinos, muons and electrons/positrons; muons in turn decay into neutrinos and electrons/positrons. Electron/positron pairs are also directly injected in proton-photon collisions. For a complete list of expressions regarding the injection functions of secondary particles, see  Reynoso \& Romero (2009), Romero \& Vila (2008) and references therein. 

The isotropic steady-state energy distribution of relativistic primary and secondary particles in the jet comoving reference frame, $N(E,z)$ (erg$^{-1}$\,cm$^{-3}$), is calculated solving the transport equation (Khangulyan et al. 2008)

\begin{equation}
v_{\rm{conv}}\frac{\partial N}{\partial z} + \frac{\partial}{\partial E}\left(\frac{dE}{dt}N\right)+ \frac{N}{T_{\rm{dec}}} = Q(E,z). 
\label{eq:transport_eq}
\end{equation}

\vspace{0.2cm} 

\noindent This equation takes into account particle injection, convection, energy losses and removal of particles due to decay. The convective term $v_{\rm{conv}}\partial N/\partial z$, where $v_{\rm{conv}}\sim v_{\rm{jet}}$ is the particle convection velocity, allows to incorporate the effect of the variation with $z$ of the parameters that characterize the acceleration region. This is not accounted for in one-zone models where the acceleration/emission region is assumed to be homogeneous, or at least thin enough to neglect all spatial dependence. The decay term is non-zero only for pions and muons.  

\noindent We include both adiabatic and radiative energy losses. In the case of leptons (primary electrons, secondary electron/positron pairs and muons) we consider losses due to synchrotron radiation, relativistic Bremsstrahlung and inverse Compton scattering. For protons and pions we take into account cooling due to synchrotron radiation and inelastic collisions with photons and thermal protons. The synchrotron field of primary electrons is used as target photon field for inverse Compton scattering, proton-photon and pion-photon collisions. Convenient expressions for the energy loss rates for all these processes can be found in Reynoso \& Romero (2009) and Romero \& Vila (2008). 

The maximum energy of primary protons and electrons at fixed $z$ is determined equating the total energy loss rate  with the acceleration rate $\left.dE/dt\right|_{\rm{acc}} = \eta\,e\,c\,B(z)$. The acceleration efficiency $\eta<1$ depends on the detailed physics of the acceleration mechanism; we simply regard it as a free parameter of the model.

For a given set of values of the model parameters, we calculate the radiative output due to all the interaction processes mentioned above. The correspondent formulas for the photon emissivities and other details about the calculations are presented in Romero \& Vila (2008), Reynoso \& Romero (2009) and references therein. 

Finally, we estimate the correction to the primary emission spectrum due to internal photon absorption through pair creation, $\gamma\gamma\rightarrow e^+e^-$. The opacity $\tau_{\gamma\gamma}$ for this process is calculated as in Gould \& Schr\'eder (1966). 

\section{Results}

\begin{figure}[b]
	\centering
	\includegraphics[width = 0.325\textwidth, keepaspectratio, trim = 25 18 25 25, clip]{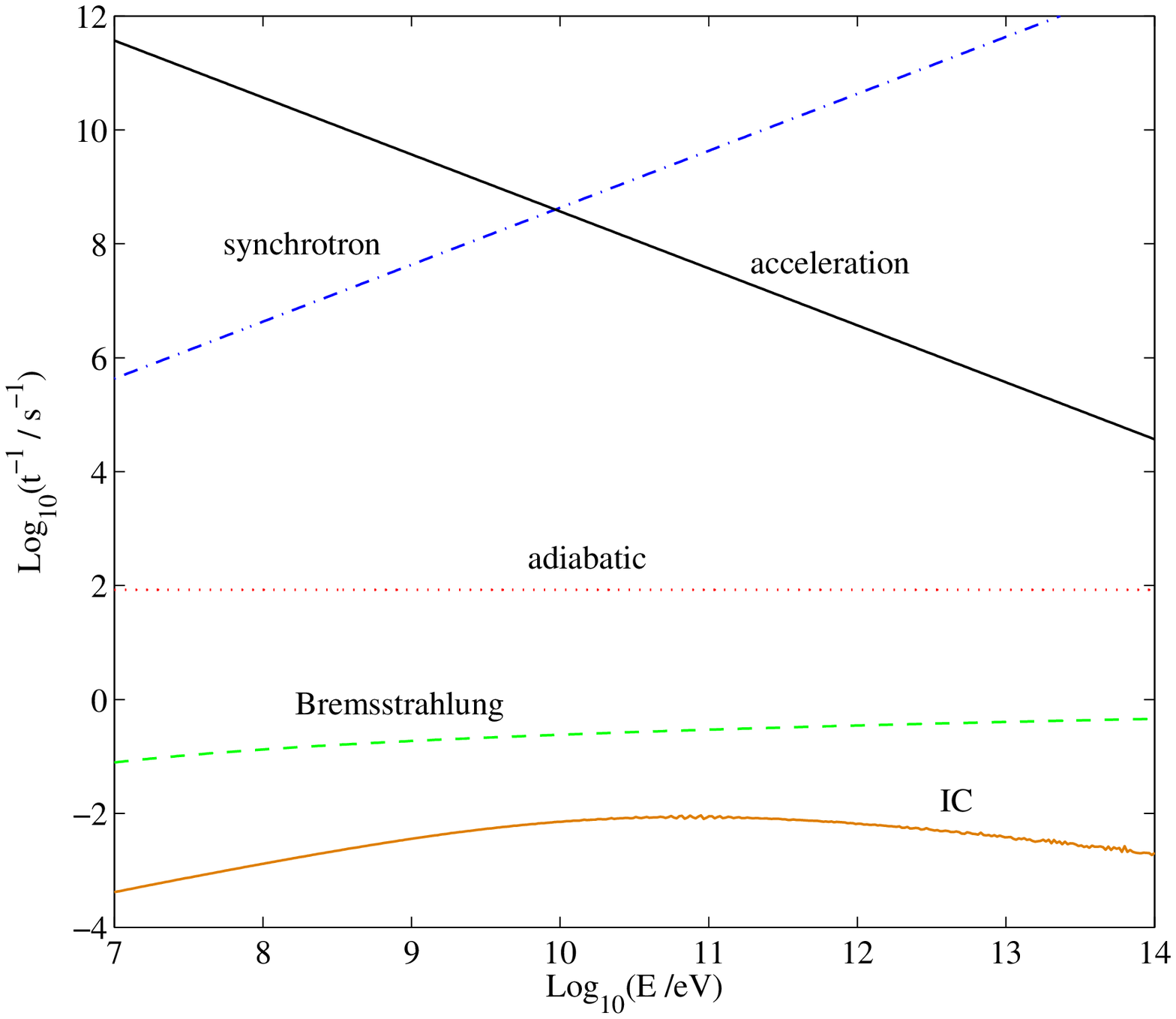}
	\includegraphics[width = 0.325\textwidth, keepaspectratio, trim = 25 18 25 25, clip]{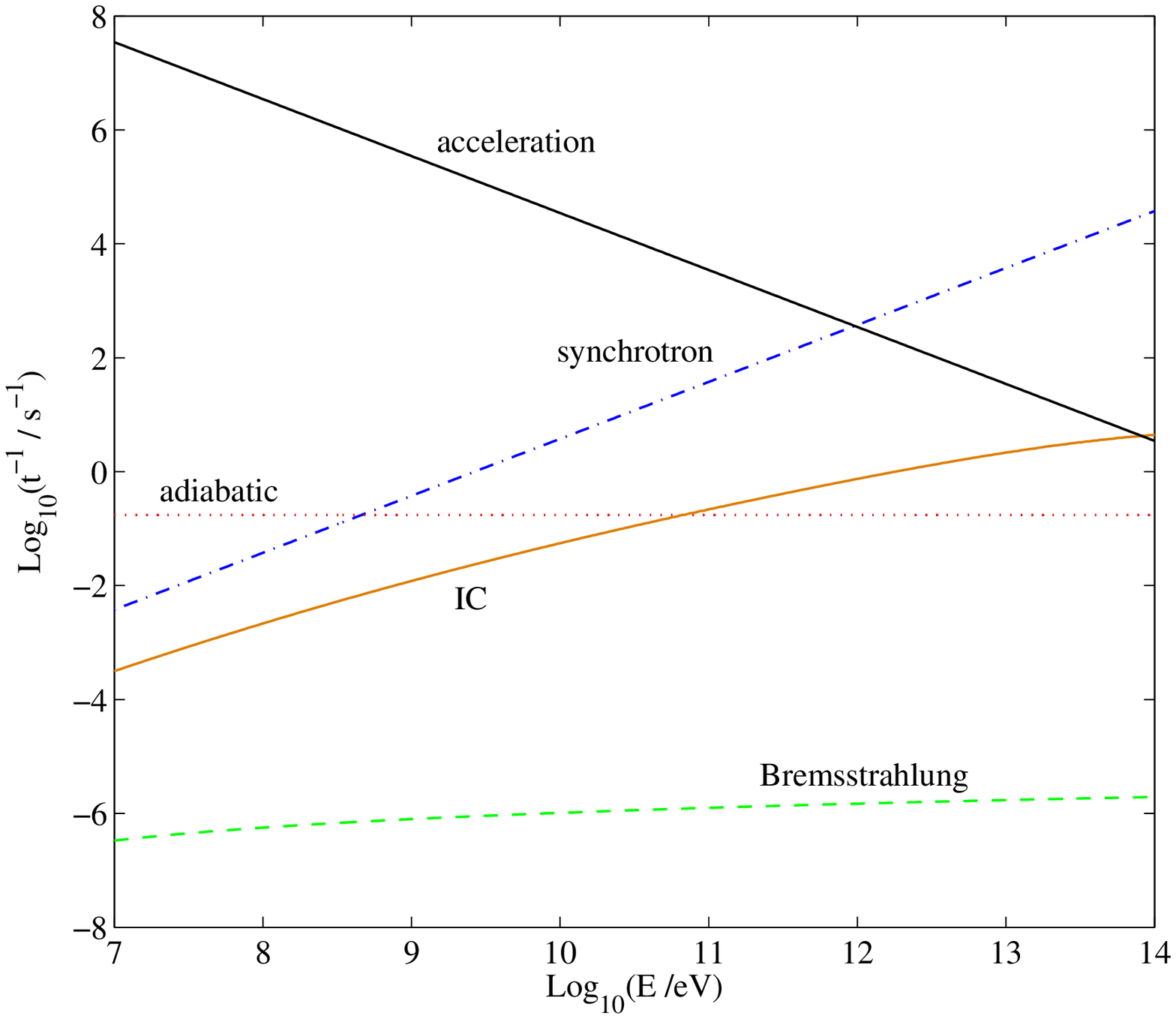}
	\includegraphics[width = 0.325\textwidth, keepaspectratio, trim = 25 18 25 25, clip]{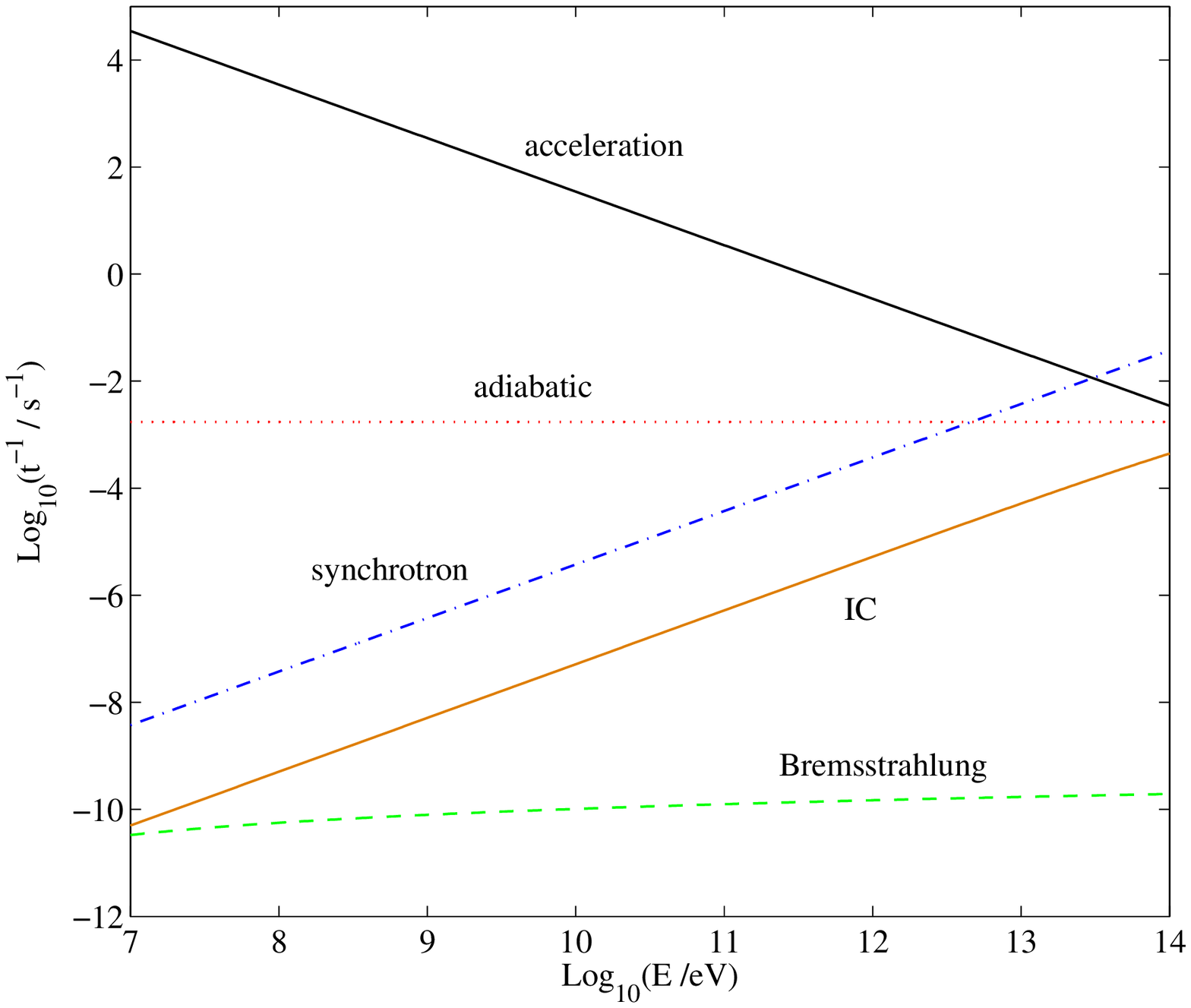}		
	\caption{Acceleration and cooling rates for primary relativistic electrons, calculated for the parameters of model A at different distances from the black hole. From left to right: $z_{\rm{acc}}$, $z_{\rm{max}}$ and $z_{\rm{end}}$.}
	\label{fig:cooling_electrons}
\end{figure}

\begin{figure}[b]
	\centering
	\includegraphics[width = 0.325\textwidth, keepaspectratio, trim = 25 18 25 25, clip]{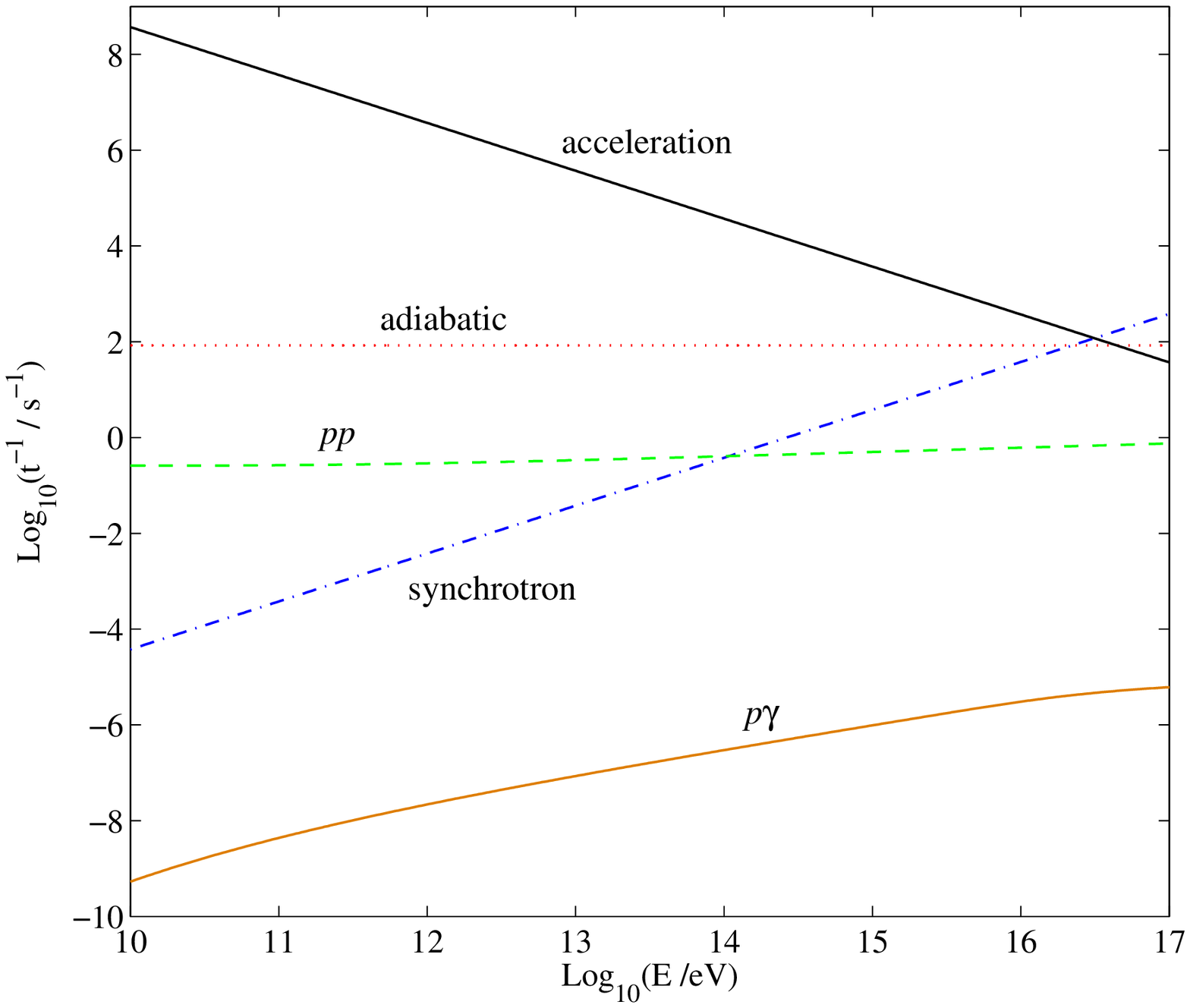}
	\includegraphics[width = 0.325\textwidth, keepaspectratio, trim = 25 18 25 25, clip]{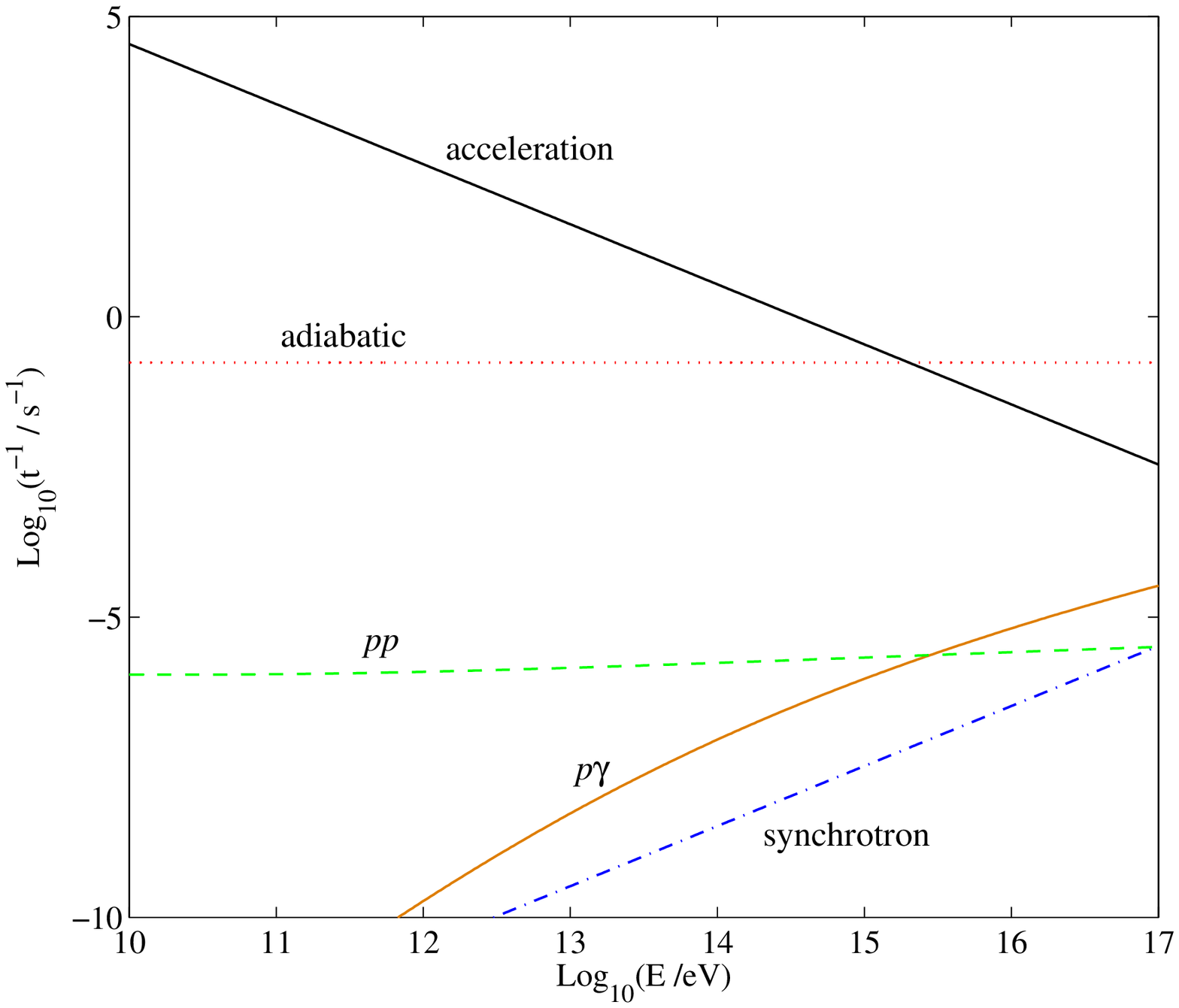}
	\includegraphics[width = 0.325\textwidth, keepaspectratio, trim = 25 18 25 25, clip]{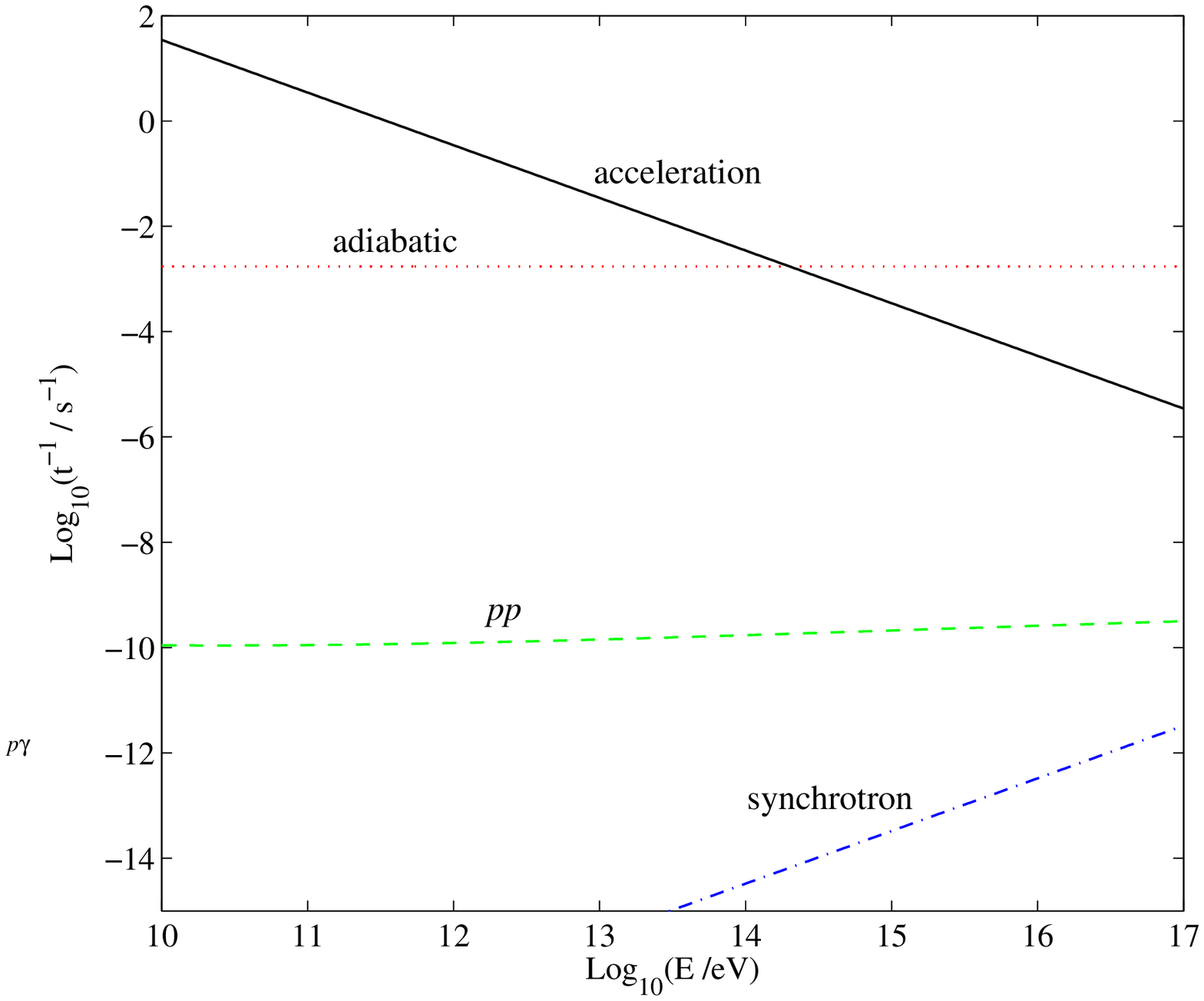}		
	\caption{Idem Figure \protect\ref{fig:cooling_electrons} for relativistic primary protons.}
	\label{fig:cooling_protons}
\end{figure}

\begin{figure}
	\centering
	\includegraphics[width = 0.4\textwidth, keepaspectratio, trim = 10 10 10 10, clip]{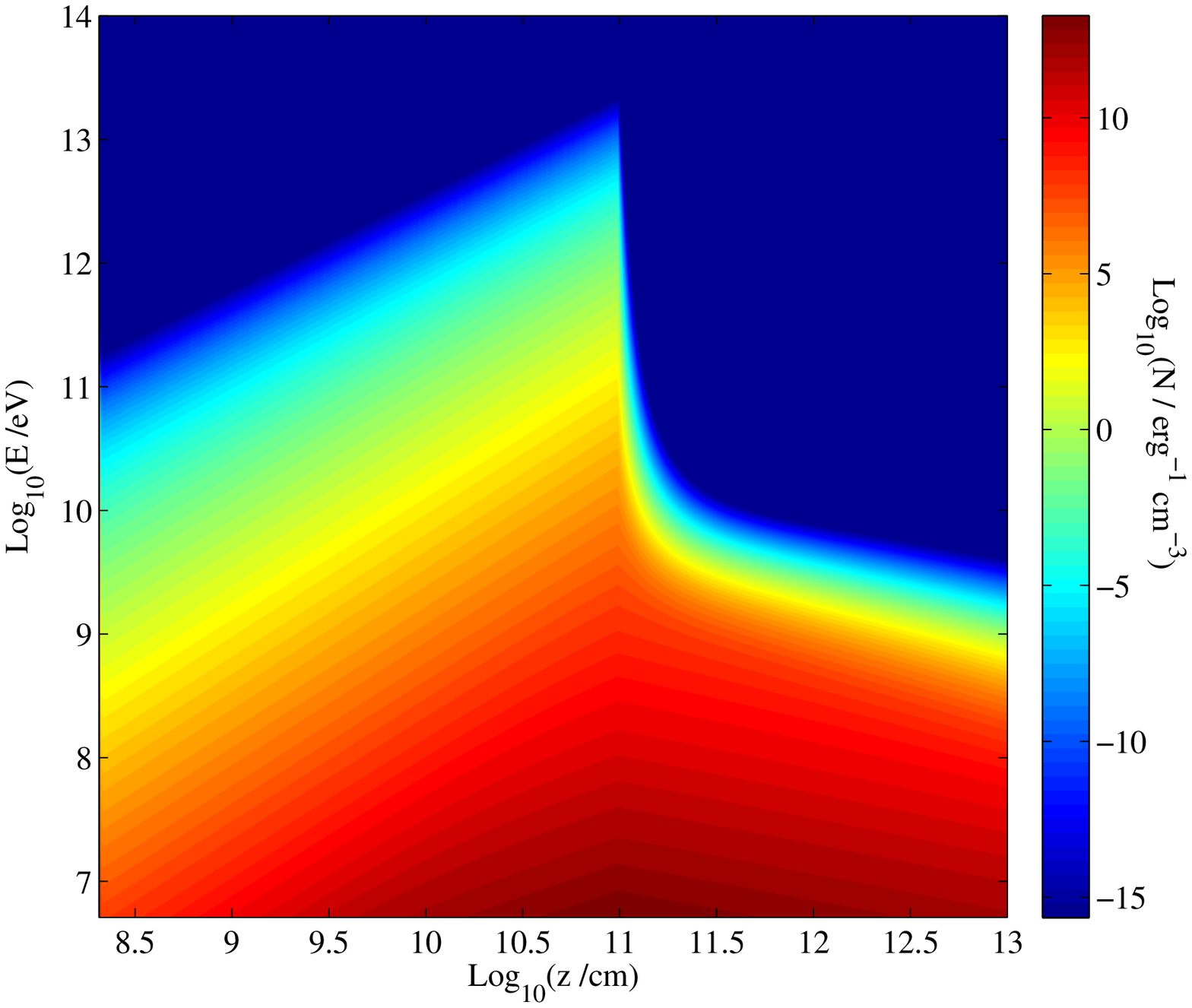}
	\includegraphics[width = 0.4\textwidth, keepaspectratio, trim = 10 10 10 10, clip]{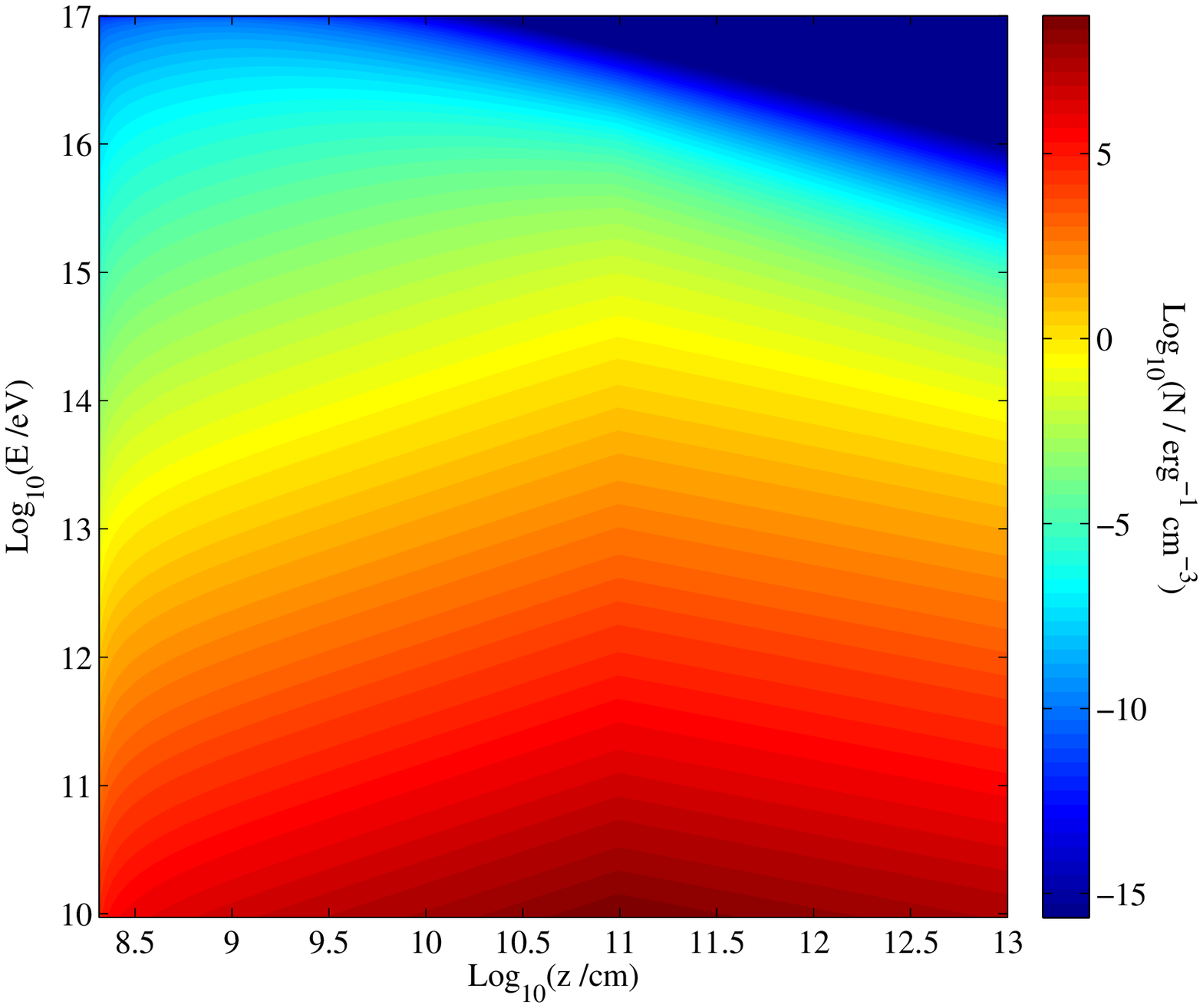}
	\caption{Distributions of relativistic primary electrons (left) and protons (right) as a function of energy and distance to the black hole, for the parameters of model A.}
	\label{fig:distributions}
\end{figure}

\begin{figure}[!h]
	\centering
	\includegraphics[width = 0.49\textwidth, keepaspectratio, trim = 15 20 70 40, clip]{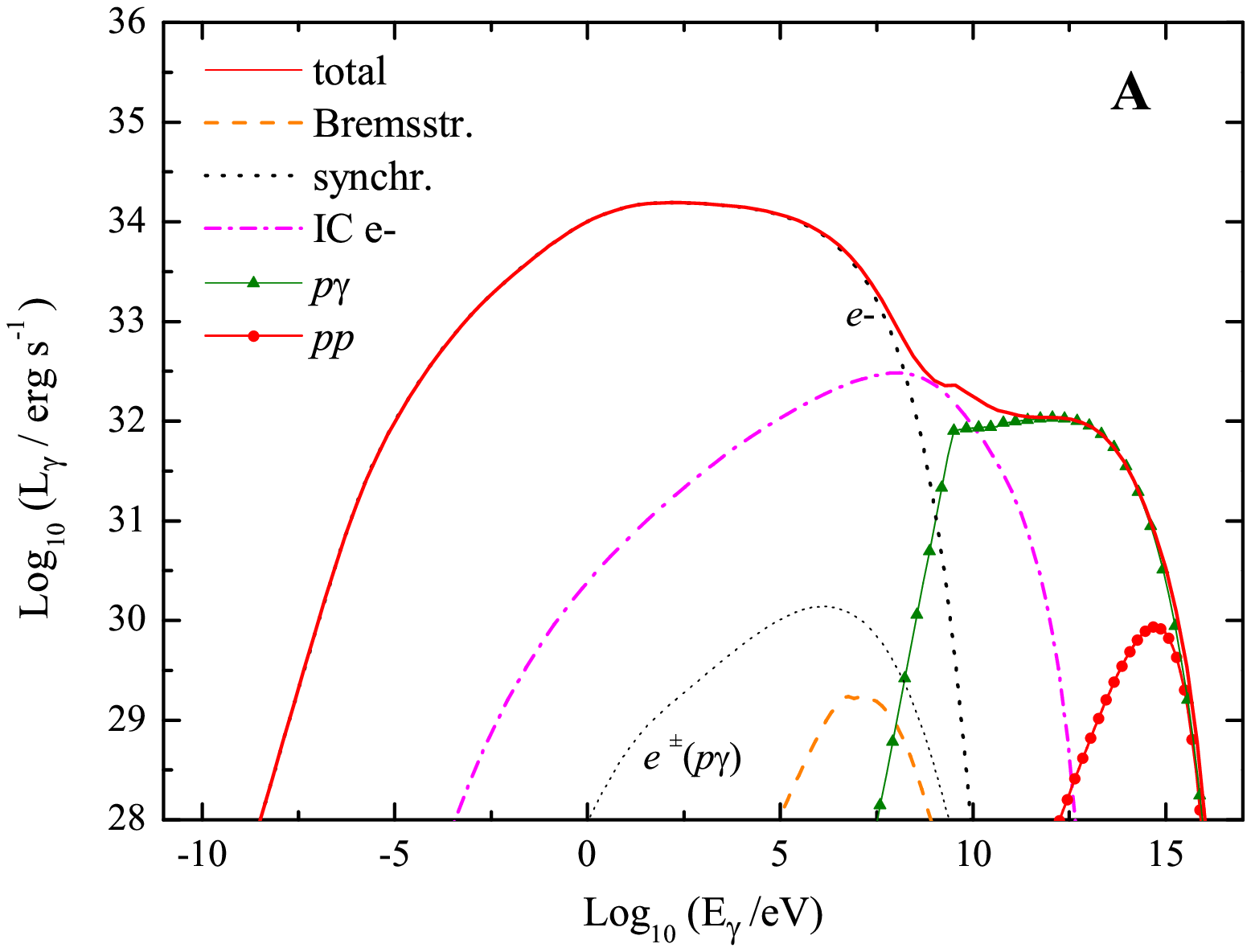}	
	\includegraphics[width = 0.49\textwidth, keepaspectratio, trim = 15 20 70 40, clip]{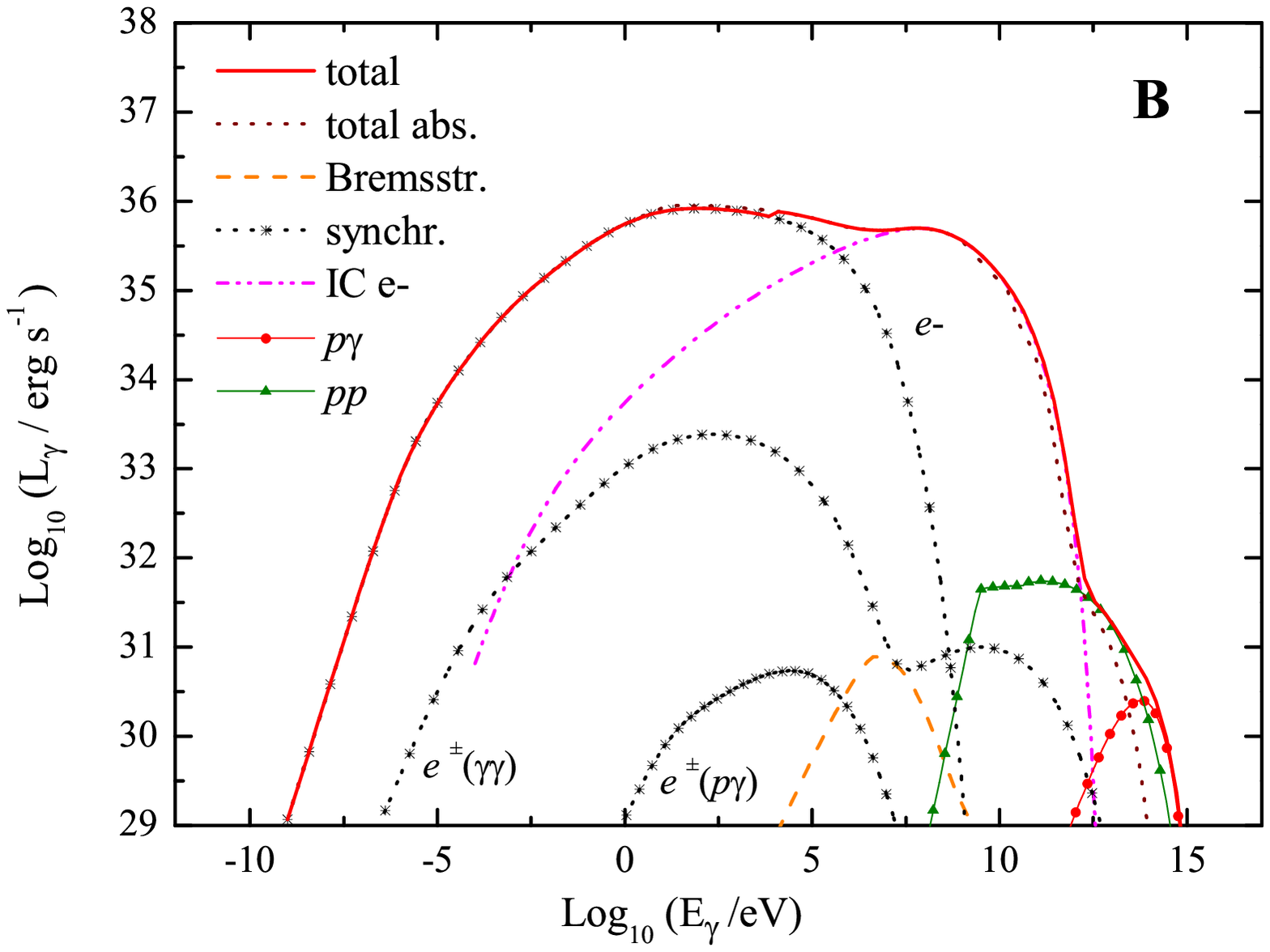}	
	\caption{Spectral energy distributions obtained for two different sets of model parameters. In the synchrotron spectra, $e^-$, $e^\pm(\gamma\gamma)$ and $e^\pm(p\gamma)$ indicate primary electrons and pairs injected through photon-photon annihilation and direct pair creation in proton-photon collisions, respectively.}
	\label{fig:SEDs}
\end{figure}

\begin{figure}[!h]
	\centering
	\includegraphics[width = 0.35\textwidth, keepaspectratio, trim = 15 5 40 20, clip]{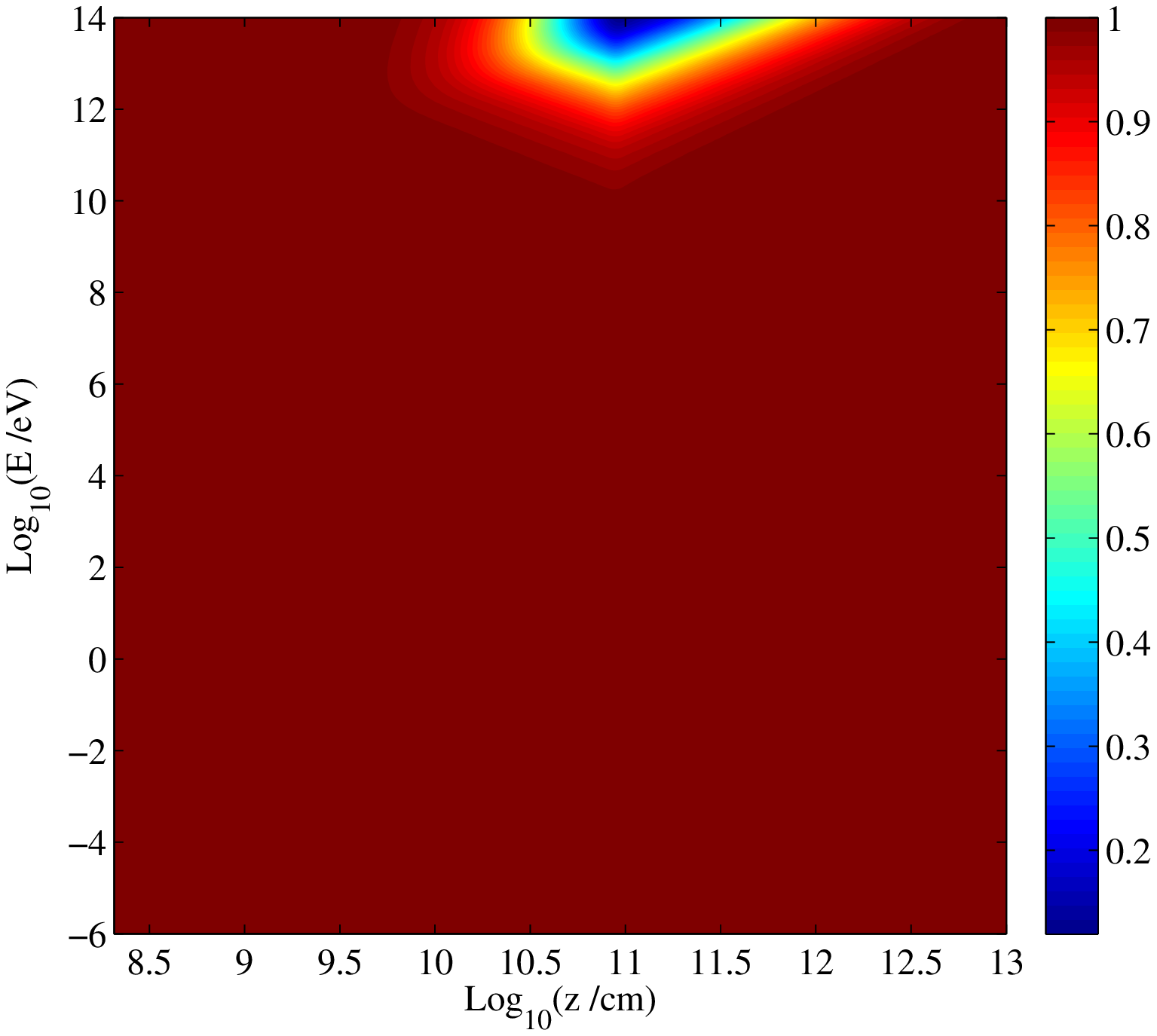}	
	\includegraphics[width = 0.365\textwidth, keepaspectratio, trim = 15 7 40 20, clip]{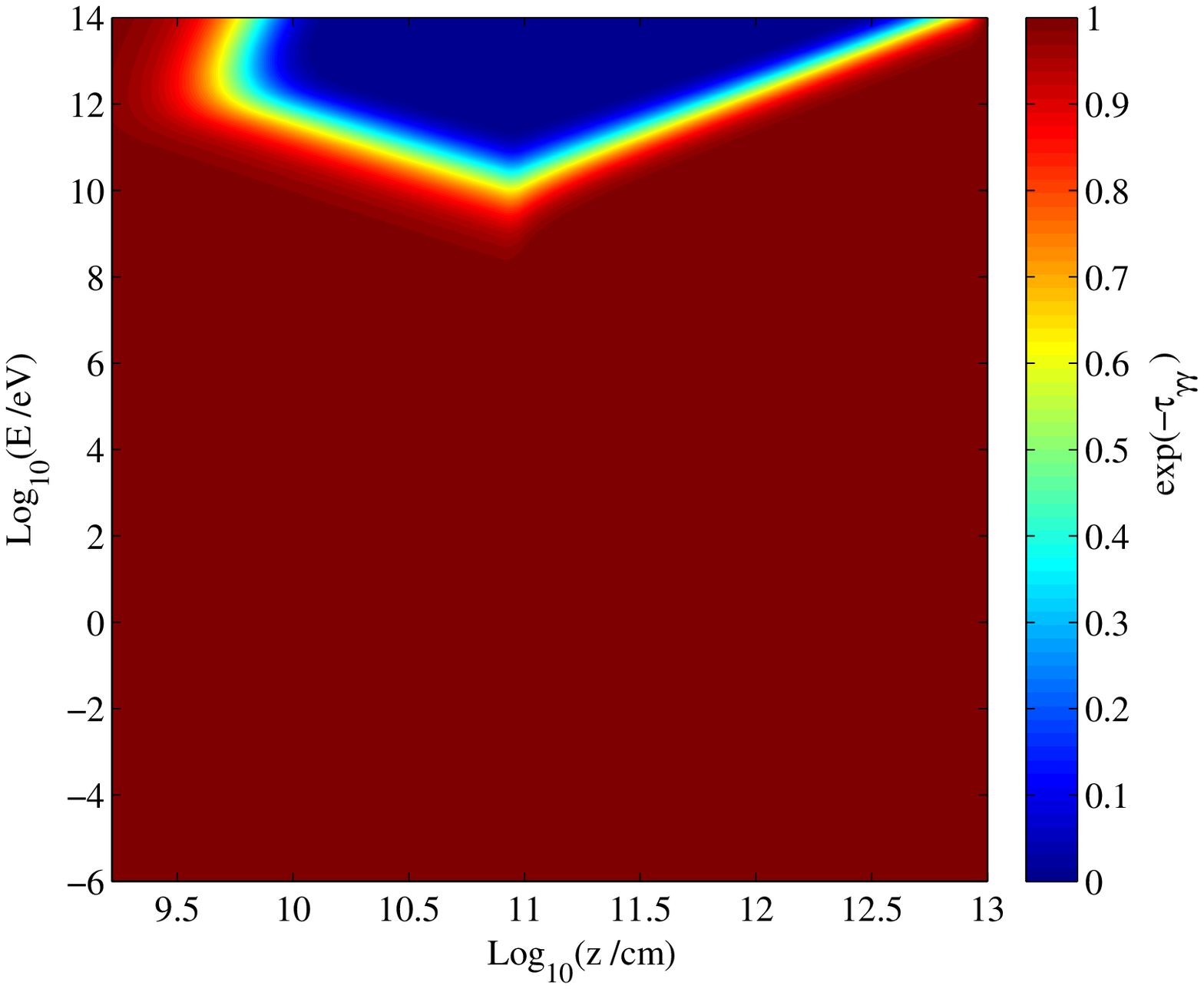}	
	\caption{Attenuation factor $\exp(-\tau_{\gamma\gamma})$ for models A (left) and B (right).}
	\label{fig:attenuation}
\end{figure}

Figures \ref{fig:cooling_electrons} to \ref{fig:attenuation} show the results obtained for two generic sets of values of the model parameters, cases A and B in Table \ref{tab:model-parameters}. The cooling rates for primary protons and electrons in model A are plotted in Figures \ref{fig:cooling_electrons} and \ref{fig:cooling_protons}. Synchrotron and adiabatic losses are the main cooling mechanisms of electrons for all $z$, whereas adiabatic losses always dominate for protons. Cooling due to those processes that involve interaction with the synchrotron photon field becomes negligible for $z>z_{\rm{max}}$. This is because in this region the number of high-energy electrons decreases drastically (see below) and consequently the synchrotron emissivity is quenched.

\begin{table}[h]
\begin{center}
\small
\caption{Values of the relevant parameters of the model.}
\label{tab:model-parameters}
\begin{tabular}{l|c|c|c|c}
\hline 
Parameter						 & Model A					  & Model B							& XTE 2000				 & XTE 2005\\
\hline
$M_{\rm{BH}}$				 & $10M_\odot$  			& $10M_\odot$  				& $8.5M_\odot$  	 & $8.5M_\odot$  \\[0.05cm]
$z_{\rm{acc}}$       & $1.6\times10^9$ cm	& $1.6\times10^9$ cm	& $3\times10^8$ cm & $6\times10^8$ cm\\[0.05cm]   
$z_{\rm{max}}$       & $10^{11}$ cm				& $10^{11}$ cm				& $3\times10^9$ cm & $6\times10^9$ cm\\ [0.05cm]
$z_{\rm{end}}$       & $10^{13}$ cm				& $10^{13}$ cm				& $10^{12}$ cm 		 & $10^{12}$ cm\\[0.05cm]   
$q_{\rm{accr}}$ 		 & $0.1$							& $0.1$								& $0.2$						 & $0.1$\\[0.05cm]               
$\eta$               & $0.2$							& $10^{-2}$						& $10^{-2}$				 & $10^{-2}$\\[0.05cm]         
$a$            		   & $100$	  					& $1$	  						& $1$							 & $5$\\[0.05cm]            
$\alpha$             & $2$	  						& $2$	  							& $1.5$						 & $1.5$\\[0.05cm]           
$\gamma_{\rm{min}}^{(*)}$  & $10$								& $10$								& $50$						 & $50$\\[0.05cm]    
\hline
\end{tabular}
\end{center}
$\qquad\qquad\qquad^{(*)}$ \footnotesize{Minimum Lorentz factor of primary protons and electrons.}
\end{table}

\normalsize

The evolution of the particle distributions for model A is shown in Figure \ref{fig:distributions}. At fixed $z$ the distributions depend on energy as a power law: $N_e\propto E_e^{-3}$ for electrons and $N_p\propto E_p^{-2}$ in the case of protons. For $z>z_{\rm{max}}$, where $Q\approx0$, there is no acceleration and particles only lose energy. The number of high-energy electrons decreases since these cool much faster than the low-energy ones. The same is true for protons, but the effect is less pronounced since proton cooling is slower.     

A number of spectral shapes can be obtained varying the value of the parameters. Figure \ref{fig:SEDs} shows the spectral energy distributions (SEDs) for the parameters of models A and B.  In case B, where $a=1$ and $\eta=10^{-2}$, synchrotron and IC radiation from primary electrons completely dominate the spectrum, reaching a maximum luminosity $L_\gamma\approx 10^{36}$ erg s$^{-1}$. The high-energy emission due to $pp$ interactions is negligible. Case A corresponds to $a=100$ and $\eta = 0.1$. The power injected in primary electrons is 100 times smaller than in the previous model, and then the synchrotron luminosity reaches only $L_\gamma\approx 10^{34}$ erg s$^{-1}$. This in turn reduces the IC emission levels since there are considerably less target photons. For $E_\gamma \gtrsim 1$ GeV the dominant contribution is due to $pp$ interaction, with a luminosity  $L_\gamma\approx 10^{32}$ erg s$^{-1}$. Radiation from secondary particles is not significant in any case. 

The attenuation factor $\exp(-\tau_{\gamma\gamma})$ as a function of photon energy and $z$ is shown in Figure \ref{fig:attenuation}. Absorption is only important for photons with $E_\gamma\gtrsim 1$ TeV, that annihilate against low-energy photons of the electron synchrotron field.  The effect is therefore stronger in model B.  The effect on the SED is, however, not significant in any case. 

\begin{figure}[!h]
	\centering
	\includegraphics[width = 0.385\textwidth, keepaspectratio, trim = 25 15 20 20, clip]{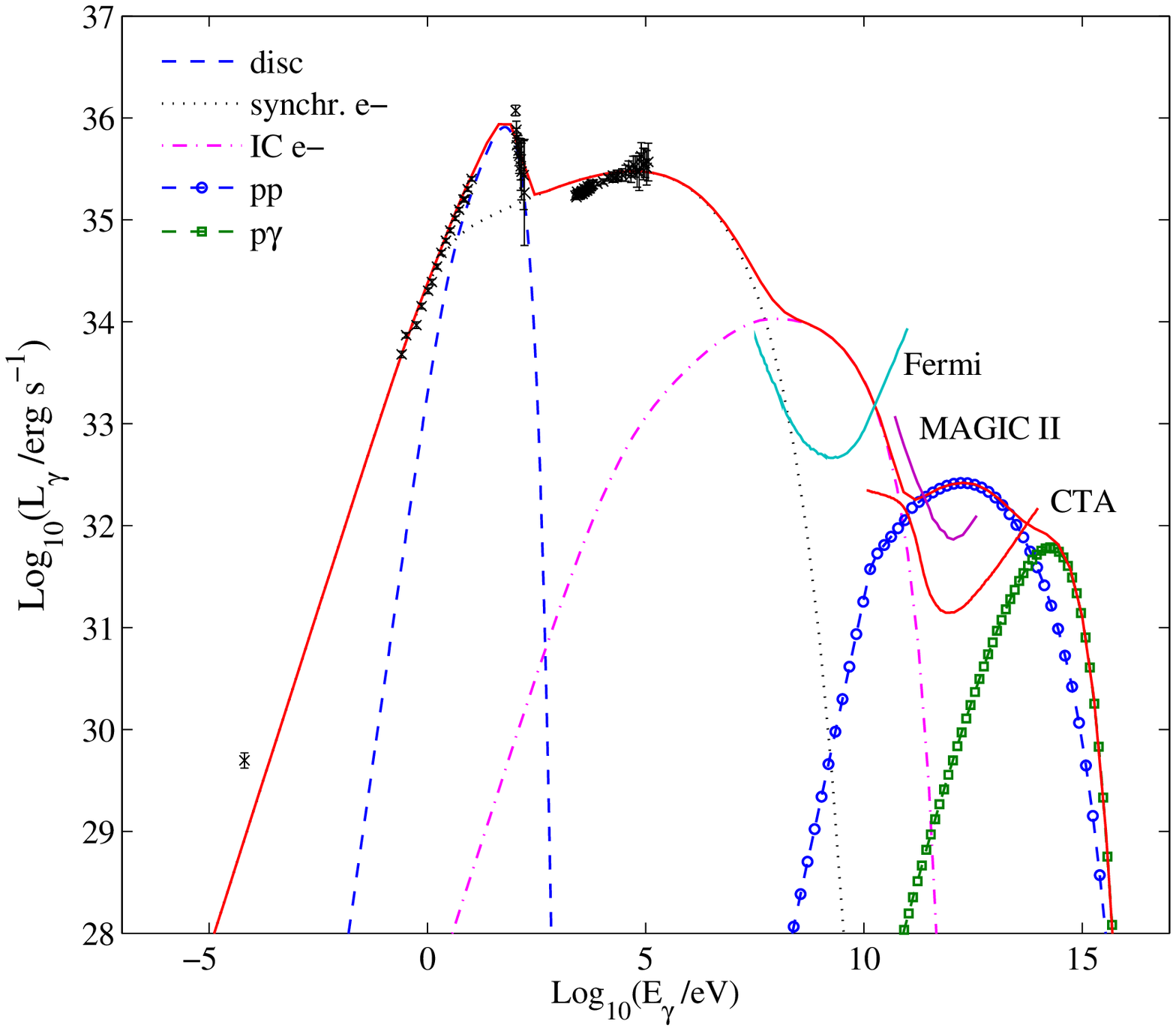}
	\includegraphics[width = 0.4\textwidth, keepaspectratio, trim = 25 15 20 20, clip]{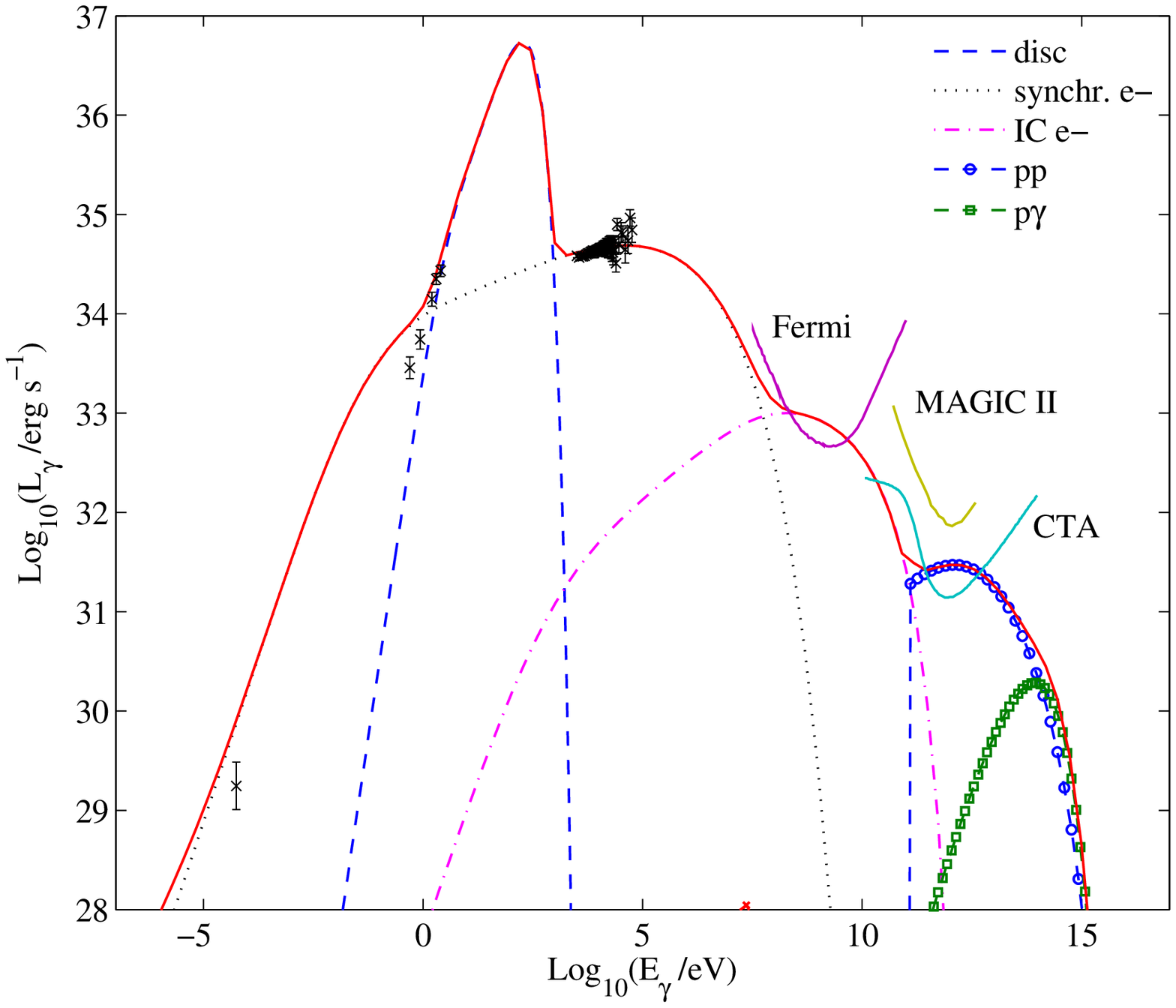}
	\caption{Fits to the observed spectrum of XTE J1118+480 in low-hard state during the outbursts of 2000 (left) and 2005 (right). }
	\label{fig:fits_XTE}
\end{figure}

We applied the model to fit the observed spectrum of the LMMQ XTE J1118+480 in the low-hard state during the outbursts of 2000 (McClintock et al. 2001) and 2005 (Zurita et al. 2006). Here we also included the  emission of a standard thin accretion disk. The best fit models were obtained for the parameters listed in Table \ref{tab:model-parameters}; the correspondent SEDs are shown in Figure \ref{fig:fits_XTE}. According to these results, the source might be marginally detected by Fermi in the GeV range during a low-luminosity low-hard state such as the case of the 2005 outburst. However, during a more luminous low-hard state as in 2000, also the hadronic contribution to the emission spectrum at TeV energies might be detected with MAGIC II and the future array CTA. 

\section{Conclusions}

We have developed a model for the broadband spectrum of jets in LMMQs that includes the radiative contribution of electrons, protons and secondary particles generated by several interaction processes. The energy distributions of all species were obtained solving the transport equation in an inhomogeneous region, taking into account cooling, injection, convection and decay. We also assessed the effect of photon absorption in the jet. 
We applied the model to fit the  observed spectrum of the low-mass X-ray binary XTE J1118+480 and make predictions for its high-energy emission. According to our results, luminous low-hard state outbursts of this source might be detected at GeV and TeV energies with current and future gamma-ray instruments.  

\acknowledgments

G.S.V. thanks Iva \v{S}nidari\'c  for the data on the sensitivity of MAGIC II. This work was supported by CONICET grant PIP 0078, by and the Spanish Ministerio de Ciencia e Innovaci\'on (MINCINN) under grant AYA 2010-21782-C03-01.

\end{document}